S. L. Cherkas[1], V. L. Kalashnikov[2]

[1]*Institute for Nuclear Problems, Belarus State University, Minsk, Belarus*
[2]*Institute of Photonics, Vienna University of Technology, Vienna, Austria*

# THEORY OF GRAVITY ADMITTING ARBITRARY CHOICE OF THE ENERGY DENSITY REFERENCE LEVEL

Five-vectors theory of gravity is proposed, which admits an arbitrary choice of the energy density reference level. This theory is formulated as the constraint theory, where the Lagrange multipliers turn out to be restricted to some class of vector fields unlike the General Relativity (GR), where they are arbitrary. The possible cosmological implication of the model proposed is that the residual vacuum fluctuations dominate during the whole evolution of the universe. That resembles the universe having a nearly linear dependence of scale factor on cosmic time.

General relativity (GR) is the elegant theory of geometrization of physical laws based on unification of space and time. It manifests striking stability relatively different modifications [1,2]. On the other hand, a long experience faces the challenges of the space-time quantization in terms of GR [3-7]. One may assume that the difficulties with quantization could originate namely as a result of some redundant rigidity of GR. It is interesting that modifications of GR such as string theory [8], loop quantum gravity [9] and others alternatives (see, e.g. [10]) follow this way of maximal symmetrization of underlying space-time laws.

Still, it seems reasonable to probe some alternatives which break the unity of space and time. Along the way, possibilities are abundant. To obtain a physically meaningful result, one must not dissent too far from GR. The examples of such theories disregarding a unity of space-time are [11-16], where the power counting renormalizability [12] or the conformal invariance [13] are reached. Another example is the unimodular gravity, in which the cosmological constant could be redefined (see [15] and references therein). In the present paper, we propose the five-vectors theory (FVT) of gravity admitting an arbitrary choice of the energy density reference level. That allows omitting a main part of the vacuum energy, whereas residual vacuum fluctuations dominate during whole universe evolution and convert the latter into the Milne-type universe [11,16], where the scale factor grows linearly with cosmic time. Original Milne universe is empty and has negative curvature, whereas FVT predicts the flat and not empty universe with the accelerated expansion in the vicinity of small redshifts.

In the first section of the paper, GR in the conformal time gauge will be preliminarily considered. The modification of GR in this particular gauge will be obtained in the second section. The Schwarzschild solution within the frameworks of FVT will be considered in the third section. The fourth section will demonstrate how the problem of the primary divergence $M_p^4$ in the vacuum energy density could be solved. Besides, the universe deceleration parameter is discussed.

**1. GR in the conformal time gauge.** The action of a system including one scalar field $\phi$ and a point particle with the rest-mass $m_0$ has the form [17]:

$$S = -\frac{M_p^2}{12}\int(\mathcal{G}+2\Lambda)\sqrt{-g}\,d^4x + \frac{1}{2}\int(\partial_\mu\phi g^{\mu\nu}\partial_\nu\phi - m^2\phi^2)\sqrt{-g}\,d^4x - m_0\int\sqrt{-g_{\mu\nu}\frac{dX^\mu}{d\eta}\frac{dX^\nu}{d\eta}}\,d\eta, \quad (1)$$

where $\mathcal{G} = g^{\alpha\beta}\left(\Gamma^\rho_{\alpha\nu}\Gamma^\nu_{\beta\rho} - \Gamma^\nu_{\alpha\beta}\Gamma^\rho_{\nu\rho}\right)$, $M_p$ is the Planck mass, which is chosen as $M_p = \sqrt{\frac{3}{4\pi G}}$, and $\Lambda$ is the cosmological constant, which is included with illustrative purposes. Here, and everywhere below zero variations of dynamical variables on a boundary are assumed.

Let us write an interval in the ADM form [18]

$$ds^2 = a^2 N^2 d\eta^2 - \gamma_{ij}(dx^i + N^i d\eta)(dx^j + N^j d\eta),$$

where $\gamma_{ij}$ is the induced three metric denoted by latin indexes, $a = \gamma^{1/6}$ is the local scale factor, $\gamma = \det\gamma_{ij}$. Thus, $\eta$ is the conformal time if $N$ equals unity. Up to the total spatial derivative term, the action (1) becomes

$$S = \int N\left(\frac{M_p^2}{12}a^4(K_{ij}K^{ij} - K^2 + R^{(3)} - 2\Lambda) + \frac{a^2(\partial_\eta\phi)^2}{2N^2} - \frac{a^4 m^2\phi^2}{2} - \frac{a^2}{N^2}N^i\partial_i\phi\partial_\eta\phi\right.$$
$$\left. + \frac{a^2}{2N^2}N^i\partial_i\phi N^j\partial_j\phi - m_0\sqrt{a^2 - \frac{1}{N^2}\gamma_{ij}\left(\frac{dX^i}{d\eta}+N^i\right)\left(\frac{dX^j}{d\eta}+N^j\right)}\delta^{(3)}(\mathbf{X}(\eta)-\mathbf{x})\right)d^3x d\eta, \quad (2)$$

where $K_{ij} = \frac{1}{2aN}(D_j N_i + D_i N_j - \partial_\eta\gamma_{ij})$ and $K = \gamma^{ij}K_{ij}$ [3]. Covariant derivatives $D_j$ and rising indexes are taken using the metric $\gamma^{ij}$. Let us set $M_p = 1$ for the simplicity of the intermediate calculations.

For ``hamiltonization'' of the theory, the momentums are needed:

$$\pi^{ij} = \frac{\delta S}{\delta(\partial_\eta\gamma_{ij})} = -\frac{1}{12}a^3(K^{ij} - \gamma^{ij}K), \quad \pi_\phi = \frac{\delta S}{\delta(\partial_\eta\phi)} = \frac{a^2}{N}(\partial_\eta\phi - N^i\partial_i\phi),$$

$$P_i(\eta) = \frac{\delta S}{\delta\left(\frac{dX^i}{d\eta}\right)} = \frac{m_0\gamma_{ij}\left(\frac{dX^j}{d\eta}+N^j\right)}{\sqrt{a^2 N^2 - \gamma_{nm}\left(\frac{dX^n}{d\eta}+N^n\right)\left(\frac{dX^m}{d\eta}+N^m\right)}}. \quad (3)$$

The momentums corresponding to $N$ and $N^i$ equal zero. Thus, the action of the system considered as the extended Hamiltonian system [19] takes the form:

$$S = \int(\pi^{ij}\partial_\eta\gamma_{ij} + \pi_\phi\partial_\eta\phi)d^3x d\eta + \int P_i(\eta)X^i(\eta)d\eta - \int H^{(1)}d\eta, \quad (4)$$

where

$$H^{(1)} = \int (N\mathcal{H} + N^i \mathcal{P}_i) d^3 x. \tag{5}$$

Variation of the action given by (4) should be taken over $\pi^{ij}(x)$, $\pi_\phi(x)$, $\gamma_{ij}(x)$, $\phi(x)$ and $P_i(\eta)$, $X^i(\eta)$ independently. The Hamiltonian $H^{(1)}$ determines evolution in an arbitrary gauge [19,20]. Besides the equations of motion, the constraints $\mathcal{H} = 0$ and $\mathcal{P}_i = 0$ arise by variation of the action (4), (5) over $N$ and $N^i$. The constraints are expressed through $\pi^{ij}$ and momentums as follows

$$\mathcal{H} = -\frac{\delta S}{\delta N} = 6a^{-2}(\gamma_{ik}\gamma_{jl} + \gamma_{il}\gamma_{jk} - \gamma_{ij}\gamma_{kl})\pi^{ij}\pi^{kl} - \frac{1}{12}a^4(R^{(3)} - 2\Lambda) +$$
$$\frac{\pi_\phi^2}{2a^2} + \frac{a^4 m^2 \phi^2}{2} + a\sqrt{P_i(\eta)P_j(\eta)\gamma^{ij} + m_0^2}\,\delta^{(3)}(\mathbf{X}(\eta) - \mathbf{x}), \tag{6}$$

$$\mathcal{P}_i = -\frac{\delta S}{\delta N^i} = -2\gamma_{ik}D_j\pi^{kj} + \pi_\phi \partial_i \phi - P_i(\eta)\delta^{(3)}(\mathbf{X}(\eta) - \mathbf{x}).$$

A time derivative of some quantity is given by [19]

$$\partial_\eta A = \{H^{(1)}, A\}, \tag{7}$$

where the Poisson brackets are

$$\{A, B\} = \int \left( \frac{\delta A}{\delta \pi^{ij}(x)} \frac{\delta B}{\delta \gamma_{ij}(x)} - \frac{\delta A}{\delta \gamma_{ij}(x)} \frac{\delta B}{\delta \pi^{ij}(x)} + \frac{\delta A}{\delta \pi_\phi(x)} \frac{\delta B}{\delta \phi(x)} - \frac{\delta A}{\delta \phi(x)} \frac{\delta B}{\delta \pi_\phi(x)} \right) d^3x + $$
$$\frac{\partial A}{\partial P_i} \frac{\partial B}{\partial X^i} - \frac{\partial A}{\partial X^i} \frac{\partial B}{\partial P_i}. \tag{8}$$

In particular

$$\{\pi^{ij}(x), \gamma_{kl}(x')\} = \frac{1}{2}(\delta_k^i \delta_l^j + \delta_i^l \delta_j^k)\delta^{(3)}(\mathbf{x} - \mathbf{x}'), \tag{9}$$

where $x = \{\eta, \mathbf{x}\}$, $x' = \{\eta, \mathbf{x}'\}$ and $\delta^{(3)}(\mathbf{x} - \mathbf{x}')$ is the Dirac delta function. Let us write the constraint algebra:

$$\{\mathcal{H}(x), \mathcal{H}(x')\} = -(\mathcal{P}_i(x)\tilde{\gamma}^{ij}(x) + \mathcal{P}_i(x')\tilde{\gamma}^{ij}(x'))\partial_j \delta^{(3)}(\mathbf{x}' - \mathbf{x}),$$
$$\{\mathcal{P}_i(x), \mathcal{P}_j(x')\} = -\left(\mathcal{P}_i(x)\partial_j + \mathcal{P}_j(x')\partial_i\right)\delta^{(3)}(\mathbf{x}' - \mathbf{x}), \tag{10}$$
$$\{\mathcal{H}(x), \mathcal{P}_i(x')\} = -\frac{2}{3}(\mathcal{H}(x) + \mathcal{H}(x'))\partial_i \delta^{(3)}(\mathbf{x}' - \mathbf{x}) - \frac{1}{3}\delta^{(3)}(\mathbf{x}' - \mathbf{x})\partial_i \mathcal{H}(x),$$

where $\tilde{\gamma}_{ij} = \gamma_{ij}/a^2$ is the metric with $\det \tilde{\gamma}_{ij} = 1$ and $\tilde{\gamma}^{ij}$ is its inverse metric. For the particular case of the Bianchi model, the corresponding algebra is given in [21].

One can see that the right-hand sides of Eqs. (10) turn to zero on the shell of the constraints $\mathcal{H}(x) = 0$ and $\mathcal{P}(x) = 0$. That is this system is of the first kind in terms of the theory of constraint systems [19,20]. Using the constraint algebra (10) and Eqs. (5), (7), we can calculate the evolution of constraints. Let us write it in the particular gauge $N = 1$, $N^i = 0$:

$$\partial_\eta \mathcal{H} = \partial_i \left( \tilde{\gamma}^{ij} \mathcal{P}_j \right), \tag{11}$$

$$\partial_\eta \mathcal{P}_i = \frac{1}{3}\partial_i \mathcal{H}. \tag{12}$$

Derivatives in Eqs. (10), (11), (12) are conventional partial, i.e. noncovariant derivatives. Eqs. (11),(12) agree with the result of [22] (Eqs. (5.13), (5.14) with $\kappa_1 = 1/3$, $\kappa_2 = \kappa_3 = 0$ and $\mathcal{H}_{[22]} = -12a^{-4}\mathcal{H}$, $\mathcal{P}_{[22]} = 6a^{-3}\mathcal{P}$). One has to note that the evolution of constraints governed by (11), (12) admits adding some constant to $\mathcal{H}$. This fact allows constructing a self-consistent constraints' theory admitting an arbitrary choice of the energy density reference level.

**2. Five-vectors theory of gravity.** As was shown in the previous section, the GR equations of motion admit a wider surface of the constraints than that of GR itself. That requires the conformal time gauge. In other gauges, the system moving on this wider surface will leave it if $\mathcal{H} \neq 0$, and thus, one should return to GR which demands $\mathcal{H}=0$. In the conformal time gauge, the system could move on the wider surface $\mathcal{H} = const$ permanently. One could expect that in some new theory admitting this wider surface of the constraints, the restrictions on the Lagrange multipliers will appear as opposed to GR, where the Lagrange multipliers are arbitrary. Below the version of such a theory will be exposed.

The theory describes an evolution with the time $\eta$ of a three-geometry defined by the metric tensor $\gamma_{ij}$. The metric tensor can be represented as a set of three triads $\gamma_{ij} = e_{ia}e_{ja}$, where index $a$ enumerates vectors of the triads $e_a$ and summation over $a$ is implied. In contrast to GR, we do not imply a united footing for the time and spatial coordinates.

Let us postulate an action of the theory in terms of the generalized Hamiltonian system as

$$S = \int (\pi^{ij}\partial_\eta \gamma_{ij} + \pi_\phi \partial_\eta \phi - \mathcal{H} - \Upsilon^i \partial_i \mathcal{H} - N^i \mathcal{P}_i)d^3\mathbf{x}d\eta, \tag{13}$$

where $\mathcal{H}$ and $\mathcal{P}_i$ are given by (6).

Thus, one has five vectors: three triad vectors $\mathbf{e}_a$, which are dynamical variables, and two vectors $\mathbf{N}$ and $\mathbf{\Upsilon}$. The latter vectors have not the corresponding momentums and, thereby, they are the Lagrange multipliers which have to be determined. Six constraints can be obtained by varying over $N^i(x)$ and $\Upsilon^i(x)$.

First, let us write the constraint algebra using (10). For simplicity it is written on the shell of the FVT constraints $\partial_i \mathcal{H}(x) = 0$ and $\mathcal{P}_i(x) = 0$:

$$\left.\begin{array}{l}\{\partial_i \mathcal{H}(x), \partial_j \mathcal{H}(x')\} = 0, \\ \{\mathcal{P}_i(x), \mathcal{P}_j(x')\} = 0, \\ \{\partial_j \mathcal{H}(x), \mathcal{P}_i(x')\} = -\dfrac{2}{3}\mathcal{H}(x)\partial_j \partial_i \delta^{(3)}(\mathbf{x'}-\mathbf{x}),\end{array}\right\} on\ shell \tag{14}$$

where $\mathcal{H} \neq 0$ is assumed, otherwise we return to GR. The full system of the constraints $\Phi_a$ consists of six constraints $\Phi_a = \{\partial_1 \mathcal{H}, \partial_2 \mathcal{H}...\mathcal{P}_2, \mathcal{P}_3\}$. They correspond to the Lagrange multipliers $\lambda_a = \{\Upsilon^1, \Upsilon^2...N^2, N^3\}$. Evolution of the system is governed by the Hamiltonian

$$H^{(1)} = H + \int (N^i \mathcal{P}_i + \Upsilon^i \partial_i \mathcal{H})d^3\mathbf{x}, \tag{15}$$

where $H$ is given by $H = \int \mathcal{H} d^3 x$. Restrictions on the Lagrange multipliers arise because the system should remain on the shell of constraints during evolution [19,20]:

$$\partial_\eta \Phi_a(x) = \{H, \Phi_a(x)\} + \int \{\Phi_b(x'), \Phi_a(x)\} \lambda_b(x') d^3 \mathbf{x}' = 0, \tag{16}$$

where summation over $b$ is implied. On the constraint surface $\Phi_a(x) = 0$, the matrix $M_{b,a}(x', x) = \{\Phi_b(x'), \Phi_a(x)\}$ composed of the Poisson brackets of constraints becomes

$$\mathbf{M}(x, x') = \frac{2}{3} \mathcal{H}(x) \begin{pmatrix} 0 & \nabla \otimes \nabla \\ -\nabla \otimes \nabla & 0 \end{pmatrix} \delta^{(3)}(\mathbf{x} - \mathbf{x}'), \tag{17}$$

where it is written in the form of four 3x3 blocks. Calculation of $\{H, \Phi_a(x)\} = 0$ gives zeros for all constraints $\Phi_b$, namely

$$\left. \begin{aligned} \{H, \partial_i \mathcal{H}(x)\} &= 0, \\ \{H, \mathcal{P}_i(x)\} &= \frac{1}{3} \partial_i \mathcal{H}(x) = 0. \end{aligned} \right\} \text{ on shell} \tag{18}$$

Eqs. (16), (17), (18) result in the restrictions on the Lagrange multipliers

$$\begin{pmatrix} 0 & -\nabla \otimes \nabla \\ \nabla \otimes \nabla & 0 \end{pmatrix} \begin{pmatrix} \Upsilon(x) \\ \mathbf{N}(x) \end{pmatrix} = 0, \tag{19}$$

that leads to two equations

$$\nabla (\operatorname{div} \Upsilon) = 0, \tag{20}$$

$$\nabla (\operatorname{div} \mathbf{N}) = 0. \tag{21}$$

where $\operatorname{div}$ consists of conventional partial noncovariant derivatives. Solutions of Eqs. (20), (21) are

$$\Upsilon(x) = \operatorname{rot} \mathbf{f}(x) + \mathbf{A} + a\mathbf{x} + \mathbf{B}(\mathbf{C}\mathbf{x}), \tag{22}$$

$$\mathbf{N}(x) = \operatorname{rot} \mathbf{g}(x) + \mathbf{D} + b\mathbf{x} + \mathbf{E}(\mathbf{F}\mathbf{x}), \tag{23}$$

where $\mathbf{f}$, $\mathbf{g}$ are some vector fields, $\mathbf{A}$, $\mathbf{B}$, $\mathbf{C}$, $\mathbf{D}$, $\mathbf{E}$, $\mathbf{F}$ are some vector functions of time, and $a$, $b$ are some scalar functions of time.

Now we can calculate the time evolution of $\mathcal{H}$ governed by the Hamiltonian $H^{(1)}$ (15). Calculation of the time derivative of $\mathcal{H}$ on the surface of the constraints $\partial_i \mathcal{H} = 0$, $\mathcal{P}_i = 0$ gives

$$\partial_\eta \mathcal{H} = \frac{4}{3} \mathcal{H} \operatorname{div} \mathbf{N}. \quad \text{on shell} \tag{24}$$

Thus, $\mathcal{H}$ evolves exponentially with time if $\mathbf{N}$ is time-independent. One has to note that $\mathcal{H} = C$ in the most interesting physical case, when the two last terms in Eq. (23) are omitted and $\operatorname{div} \mathbf{N} = 0$. Here the constant $C$ does not depend on spatial coordinates by virtue of $\partial_i \mathcal{H} = 0$.

Let us discuss the question distinguishing GR from linear field theories, namely: could the equations of motion for the point particles be deduced from the field equations [23-27]? The equations of the FVT theory are weaker than those of GR, and do not allow using the Bianchi identities directly to obtain the restrictions to the matter energy-momentum tensor. However, the underlying insight is that the field equations valid in the all space except singular points which move in agreement with the field equations [23-26]. Because these singularities correspond to the point particles, the equations of motion for the latter are contained in the field equations. In FVT, Eq. (24) demonstrates that $\mathcal{H}_{grav} + \mathcal{H}_{particles} = F(\eta)$. Because the function $F(\eta)$ is the function of time only, it does not contain spatial singularities. Thus, the singularities in $\mathcal{H}_{grav}$ must coincide with those in $\mathcal{H}_{particles}$, so that the singularities contained in the field equations of FVT have to give the equations of motion for the point particles like those in GR.

FVT is formulated in the generalized Hamiltonian form (13). To formulate it in the Lagrange form (before Lagrange multipliers fixing), one should vary the action (13) over $\pi^{ij}$. That can be done by rewriting the term $S_1 = \int \Upsilon^i \partial_i \mathcal{H} d^3\mathbf{x} d\eta$ in the action (13) as $S_2 = -\int \mathcal{H} \partial_i \Upsilon^i d^3\mathbf{x} d\eta$ to avoid appearing the spatial derivatives of the momentums. That results in the equations expressing velocities through momentums. From these equations, one has to express the momentums through velocities and substitute them into (13). As a result, we come to the action given by (1) or (2), but with the lapse function $N$ replaced by $1 + \partial_i \Upsilon^i$.

The physical sense of FVT is very simple: the standard Einstein-Hilbert action is varied not over all the possible metrics, but over some restricted class of them. The vacuum energy problem demands that the GR invariance relatively the general coordinate transformations has to be violated. It was found that most of the field theories undergo a violation of symmetry which presents in theory initially [28]. Here we violate the general relativistic invariance restricting the class of the metrics over which the action is varied. Permitted class of the metrics is of the form

$$ds^2 = g_{\mu\nu} dx^\mu dx^\nu = a^2 \left(1 - \partial_m \Upsilon^m\right)^2 d\eta^2 - \gamma_{ij}(dx^i + N^i d\eta)(dx^j + N^j d\eta). \tag{25}$$

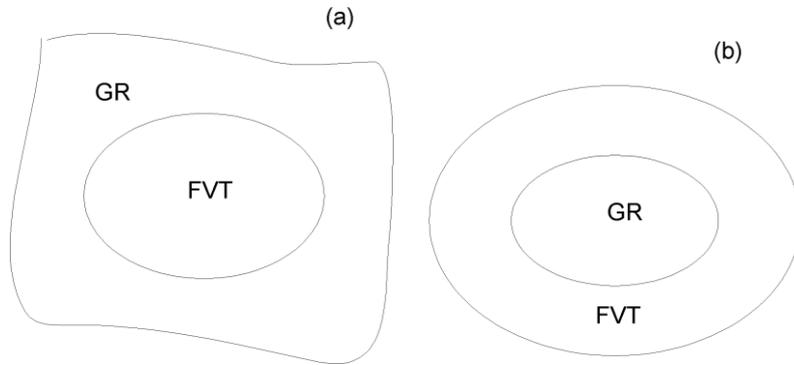

Fig. 1. Schematic comparison of GR and FVT. (a) restrictions on the Lagrange multipliers. (b) constraint surfaces in GR and FVT.

The result of the restriction is schematically represented in Fig. 1. The Lagrange multipliers in GR are arbitrary, but in FVT they are restricted by (20),(21). At the same time, the constraint surface in FVT is wider than that in GR. It should be noted that the unimodular gravity [15] also uses a mechanism of metric restriction, namely, the 4-metrics with unit determinant are considered. That violates the gauge symmetry, as well.

The expression for the gravity Lagrangian density can be rewritten in terms of triads as

$$\mathcal{L}_{grav}(x) = \frac{1}{2(1+\partial_i \Upsilon^i)} \left( \frac{1}{6} a^2 v_{bc}^2 + \frac{2}{3} a(\partial_\eta a)(\partial_i N^i) - \frac{1}{9} a^2 (\partial_i N^i)^2 - \frac{2}{3} a(N^i \partial_i a)(\partial_j N^j) + \right.$$
$$\left. 3(\partial_\eta a)(N^i \partial_i a) - \frac{3}{2}(N^i \partial_i a)^2 - \frac{3}{2}(\partial_\eta a)^2 \right) + \frac{1}{12} a^4 (1+\partial_i \Upsilon^i) R^{(3)}, \tag{26}$$

where $R^{(3)}$ is the three-dimensional curvature, which can be expressed in terms of triads:

$$R^{(3)} = -e_a^i \partial_i \gamma_{abb} + e_b^i \partial_i \gamma_{aba} + \gamma_{abf}(\gamma_{fab} - \gamma_{fba}) - \gamma_{afb} \gamma_{fba} + \gamma_{afa} \gamma_{fbb}, \tag{27}$$

where $\gamma_{abc} = \frac{1}{2}(\lambda_{abc} + \lambda_{bca} - \lambda_{cab})$ and $\lambda_{abc} = e_b^i e_c^j \partial_j e_{ai} - e_c^i e_b^j \partial_j e_{ai}$ [17]. The summation is implied both over the index $i$ of the coordinate system and over the numbers of the triad vectors $a, b...$. The quantity $v_{bc}$ is expressed as

$$v_{bc} = \frac{1}{2}\left( e_b^i v_{ci} + e_c^i v_{bi} \right), \tag{28}$$

where

$$v_{ai} = \partial_\eta e_{ai} + \frac{1}{3} e_{ai} \partial_j N^j - e_{aj} \partial_i N^j - N^j \partial_i e_{aj}. \tag{29}$$

Let us count degrees of freedom in FVT. There are five 3-vectors, i.e., fifteen quantities. However, the orientation of the triads, giving the same metrics, could be chosen arbitrary and determined by the three Euler angles. Thus, twelve degrees of freedom remains. Lagrange multipliers are restricted by (22), (23), and their remaining parts are set arbitrary. As a result, six quantities remain. Variation by $\mathbf{N}$ gives the momentum constraint, so the number of the degrees of freedom diminishes to three. Variation by $\Upsilon$ according to (21), (24) gives the constraint $\mathcal{H} = F(\eta)$, where the function of conformal time $F(\eta) \sim \exp\left( \frac{4}{3} \int \mathrm{div}\, \mathbf{N} d\eta \right)$. Thus, there are two degrees of freedom as in GR.

It may be convenient to decompose the spatial metric into the scale factor and the metric with unit determinant $\gamma_{ij} = a^2 \tilde{\gamma}_{ij} = a^2 \tilde{e}_{bi} \tilde{e}_{bj}$ (see, e.g. [29]). In this case, the Lagrangian density takes the form

$$\mathcal{L}_{grav}(x) = \frac{1}{2(1+\partial_i \Upsilon^i)} \left( -(\partial_\eta a - \frac{1}{3} a \partial_i N^i - N^i \partial_i a)^2 + \frac{1}{6} a^2 \tilde{v}_{cb}^2 \right) +$$
$$\frac{(1+\partial_i \Upsilon^i)}{12} \left( a^2 \tilde{R}^{(3)} - 4a \partial_i a \partial_j (\tilde{e}_c^j \tilde{e}_c^i) - 4a \tilde{e}_c^i \tilde{e}_c^j \partial_i \partial_j a + 2 \tilde{e}_c^i \tilde{e}_c^j \partial_i a \partial_j a \right), \tag{30}$$

where the expressions for $\tilde{v}_{bc}$, and $\tilde{R}^{(3)}$ are the same as those for $v_{bc}$, and $R^{(3)}$, but are built from the triads $\tilde{e}_{aj} = e_{aj}/a$ with the unit determinant. The indexes $i, j...$ should be risen by the matrix $\tilde{\gamma}^{ij}$, i.e. $\tilde{e}_a^i = \tilde{e}_{aj} \tilde{\gamma}^{ij}$, where $\tilde{\gamma}^{ij}$ is the inverse $\tilde{\gamma}_{ij}$.

### 3. The empty universe and Schwarzschild solution

Let us consider an example of a spherically symmetric gravitational field, which includes asymptotically empty universe and embedded Schwarzschild metric. Spherically symmetric metric belonging the class of the FVT metrics is

$$ds^2 = a^2 (d\eta^2 - \tilde{\gamma}_{ij} dx^i dx^j) = e^{2\alpha} \left( d\eta^2 - e^{-2\lambda} (d\mathbf{x})^2 - (e^{4\lambda} - e^{-2\lambda})(\mathbf{x} d\mathbf{x})^2/r^2 \right), \tag{31}$$

where $r = |\mathbf{x}|$ and $a = \exp\alpha$, $\lambda$ are functions of $\eta, r$. The isotropic coordinates used in (31) are analog of the usual Descartes coordinates. The function $\lambda$ parametrizes the metric $\tilde{\gamma}_{ij}$ with the unit determinant and reflects deflection of the conformal geometry from Euclidian one. The equations of FVT take the form

$$\mathcal{H} = e^{2\alpha}\left(-\frac{1}{2}\alpha'^2 + \frac{1}{2}\lambda'^2 - \frac{e^{2\lambda}}{6r^2} + \right.$$
$$\left. e^{-4\lambda}\left(\frac{1}{6r^2} - \frac{4}{3}\partial_r\alpha\partial_r\lambda + \frac{1}{6}\partial_r\alpha^2 + \frac{2\partial_r\alpha}{3r} + \frac{1}{3}\partial_{r,r}\alpha + \frac{7}{6}\partial_r\lambda^2 - \frac{5\partial_r\lambda}{3r} - \frac{1}{3}\partial_{r,r}\lambda\right)\right) = const, \quad (32)$$

$$\partial_r\alpha(\alpha' + 2\lambda') + \partial_r\lambda' - \partial_r\alpha' + (3/r - 3\partial_r\lambda)\lambda' = 0, \quad (33)$$

$$\alpha'' + \alpha'^2 + \lambda'^2 = e^{-4\lambda}\left(-4\partial_r\alpha\partial_r\lambda + \partial_r\alpha^2 + \frac{2\partial_r\alpha}{r} + \partial_{r,r}\alpha + \right.$$
$$\left. \frac{7}{3}\partial_r\lambda^2 - \frac{10\partial_r\lambda}{3r} - \frac{2}{3}\partial_{r,r}\lambda + \frac{1}{3r^2}\left(1 - e^{6\lambda}\right)\right), \quad (34)$$

$$\lambda'' + 2\alpha'\lambda' = \frac{2}{3}e^{-4\lambda}\left(-\partial_r\alpha\partial_r\lambda - \partial_r\alpha^2 + \partial_{r,r}\alpha + \partial_r\lambda^2 - \frac{1}{2}\partial_{r,r}\lambda - \frac{1}{r}\partial_r\alpha - \frac{1}{r}\partial_r\lambda + \frac{1}{2r^2}\left(e^{6\lambda} - 1\right)\right), \quad (35)$$

where the prime means the differentiation over $\eta$. Eq. (32) is a Hamiltonian constraint, which is satisfied up to some constant in FVT. Eq. (33) is a consequence of the momentum constraints $\mathcal{P}_i = 0$. Two other equations (34), (35) are the equations of motion. Firstly, let's consider a uniform empty flat universe with the Euclidian spatial geometry $\lambda = 0$. Eq. (32), (34) are reduced to

$$\mathcal{H} = -\frac{1}{2}\alpha'^2 e^{2\alpha} = const, \qquad \alpha'' + \alpha'^2 = 0. \quad (36)$$

An evident solution of the equations (36) is $\alpha = \ln\frac{\eta}{\eta_0}$, where $\eta_0$ is the conformal present time value at which $a = e^\alpha = 1$. Thus, the constant in this example determines the cosmological expansion. It has the unique value for all the space. Its influence on some spatially local events appears as the influence of cosmological expansion on these events. When this influence is negligible, the constant in Eq. (32) could be set to zero. Let us consider a static solution for this case, which is, certainly, a pure Schwarzschild solution. Still, it is interesting, how this pure Schwarzschild solution looks in the class of metrics (25), (31) permitted by FVT? In the static case, the time derivatives in Eqs. (32)-(35) should be set to zero. Then, one could express $\partial_{r,r}\lambda$ and $\partial_{r,r}\alpha$ from Eqs. (34), (35) and substitute them into Eq. (32). For the $const = 0$, Eq. (32) reduces to

$$-3r^2\left(\frac{d\alpha}{dr}\right)^2 + 4r\frac{d\alpha}{dr}\left(r\frac{d\lambda}{dr} - 1\right) - \left(r\frac{d\lambda}{dr} - 1\right)^2 + e^{6\lambda} = 0. \quad (37)$$

To guess the solution of (34), (35), (37) the substitution

$$\lambda = \alpha + \ln\left(\left(1 - e^{2\alpha}\right)r/r_g\right), \quad (38)$$

could be performed, where the Schwarzschild radius $r_g$ is used to make the expression under logarithm dimensionless. Finally, Eq. (37) becomes

$$r^4 e^{4\alpha}\left(e^{2\alpha}-1\right)^8 - 4\left(\frac{d\alpha}{dr}\right)^2 r_g^6 = 0, \qquad (39)$$

and allows the solution

$$\alpha(r) = \ln\left[f^{-1}\left(\frac{r^3 - r_0^3}{6r_g^3}\right)\right], \qquad (40)$$

where $f^{-1}$ is an inverse function of $f(a) = 2\ln\left(\frac{a^2}{1-a^2}\right) + \frac{30a^4 - 12a^6 - 22a^2 + 3}{6a^2\left(a^2-1\right)^3}$ and $r_0$ is constant of integration. As is shown in Fig. 2(a), the function $\alpha$ is not singular at $r=0$, but the function $\lambda$ describing conformal three geometry is singular.

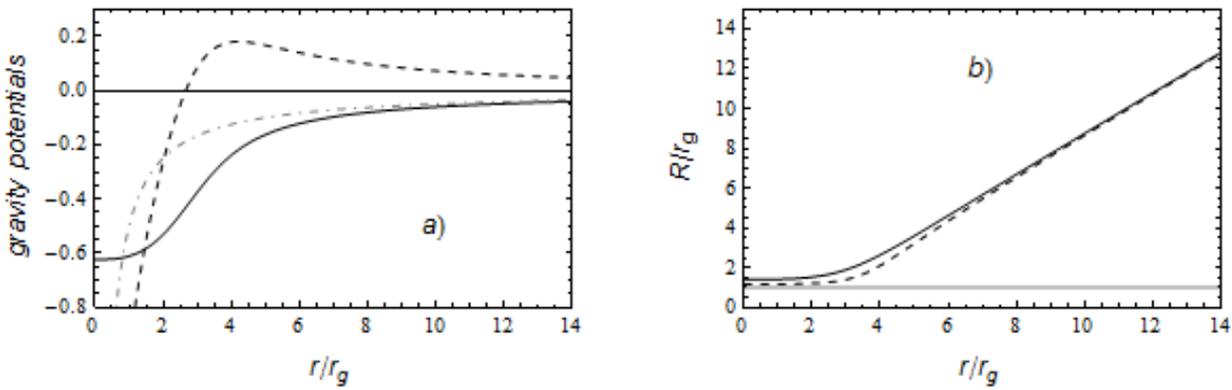

Fig. 2. (a) Parameters of the metric (31), which are referred as "gravity potentials" here. The solid line is $\alpha(r)$, the dashed line is $\lambda(r)$ and dot-dashed line is the Newton potential $\varphi = -\frac{r_g}{2r}$. (b) The plot of the coordinate transformation $R(r)$ from the interval (31), (42) to the canonical Schwarzschild one (41) for different values of the integration constant $r_0$ in (40). Solid and dashed lines correspond to $r_0 = 0$ and $r_0 = 3r_g$, respectively. Gray horizontal line corresponds to $R = r_g$.

To compare this solution with the canonical Schwarzschild one

$$ds^2 = (1 - r_g/R)dt^2 - (1 - r_g/R)^{-1}dR^2 - R^2(\sin^2\theta d\phi^2 + d\theta^2), \qquad (41)$$

let's rewrite the interval (31) in the spherical coordinates

$$ds^2 = e^{2\alpha}\left(d\eta^2 - dr^2 e^{4\lambda} + e^{-2\lambda}r^2\left(d\theta^2 + d\phi^2\sin\theta^2\right)\right). \qquad (42)$$

The coordinate transformation relates (41), (42) as $t = \eta$ and $R(r) = re^{\alpha - \lambda}$ (see Fig. 2 (b)). Regarding the canonical Schwarzschild solution, this picture corresponds to the exterior of the Schwarzschild radius, since $r = 0$ corresponds to $R > r_g$ as it is shown in Fig. 2(b). Let us give some illustrative interpretation (Fig.3) of this fact and consider the inverse transformation $r(R)$ from the canonical Schwarzschild to the FVT metric. Let we have "holed" Schwarzschild space initially. The mapping $r(R)$ could be considered as shrinking a "hole" $R \geq r_g$ to a point $r=0$, as it is shown in Fig.3. Thus, FVT repairs a "holed" Schwarzschild space-time by shrinking a hole edge into a node $r=0$ and placing a point-like particle in this node, which corresponds to the delta function term in the Lagrangian (2). The structure of static solution in FVT tell us that delta-functional sources in action have a sense only if situated in the causally reachable part of space.

It is possible to consider a test particle motion in the vicinity of $r=0$, where $\alpha \approx const$ and $\lambda = \alpha + \ln\left(\left(1-e^{2\alpha}\right)r/r_g\right) \approx const + \ln r$. Radial geodesic line obeys the equations $\ddot{\eta}=0$, $\ddot{r}+2\dot{r}^2/r=0$, where the dot means differentiation over the proper time $s$. The explicit solution is $r(\eta) = r_{in}^{2/3}\left(r_{in} - 3v(\eta-\eta_{in})\right)^{1/3}$, which implies that the particle is placed initially at $r_{in}$ and $\eta = \eta_{in}$, and has a velocity $v$ directed towards the center falls into the center $r=0$ at some finite conformal time.

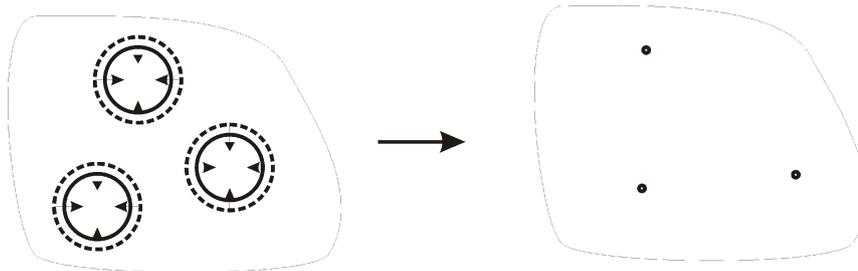

Fig. 3. Shrinking of the "holes" of the Schwarzschild radius into the nodes to place the delta-function mass sources into them.

The general case unifying these two examples can describe a model of the evolution of the spherical superclusters of $\Box\, 100\, Mps$ size in the expanding universe having $\eta_0 \Box\, 4500\, Mps$. However, it will be only an academical example because, as will be shown in the next section, the quantum vacuum should be first considered explicitly.

### 4. Domination of vacuum fluctuations in the evolution of the universe.

The possibility of an arbitrary choice of the energy reference level allows omitting the huge vacuum energy [30-34]. But the most interesting question is what remains after this omitting [33,34]? It turns out to be that the Milne-type cosmology arises as a result of residual vacuum fluctuations. To demonstrate that fact, let us consider a particular metric:

$$\gamma_{ij} = a^2(\eta)\,\mathrm{diag}\{1,1,1\}. \qquad (31)$$

Below, the scale factor $a(\eta)$ will be considered as homogenous whereas the scalar field is inhomogeneous. It should be noted, that the gravitons contribute to the vacuum energy as two minimally coupled massless scalar fields [33]. Thus, without loss of generality, the only quantum scalar field is considered here.

Both constraints and equations of motion suggest $\mathcal{H} = const$. The Hamiltonian is $H = \tilde{\mathcal{H}}(0)$, where $\tilde{\mathcal{H}}(\mathbf{k}) = \int \mathcal{H}(x)\exp(i\mathbf{k}x)d^3x$. That is the Hamiltonian $H = const$. Substitution of the Fourier representation of a scalar field $\phi(x) = \sum_{\mathbf{k}} \phi_{\mathbf{k}} \exp(i\mathbf{k}x)$ into the equation for $H$ results in

$$H = V\left(-\frac{1}{2}M_p^2 a'^2 + \frac{1}{6}M_p^2 \Lambda a^4 + \frac{1}{2}\sum_{\mathbf{k}} a^2 \phi'_\mathbf{k} \phi'_{-\mathbf{k}} + a^2 k^2 \phi_\mathbf{k} \phi_{-\mathbf{k}} + a^4 m^2 \phi_\mathbf{k} \phi_{-\mathbf{k}}\right), \tag{32}$$

where the integration $d^3x$ is performed over the normalization volume $V$, and we restored the Planck mass. Redefinitions $a^2 \to a^2/V$ and $m^2 \to m^2 V$ allow omitting the volume $V$ in intermediate calculations. The equation of motion for a scalar field is

$$\hat{\phi}''_\mathbf{k} + (k^2 + a^2 m^2)\hat{\phi}_\mathbf{k} + 2\frac{a'}{a}\hat{\phi}'_\mathbf{k} = 0. \tag{33}$$

Quantization of the scalar field [35]

$$\hat{\phi}_\mathbf{k} = \hat{a}^+_{-\mathbf{k}} \chi^*_k(\eta) + \hat{a}_\mathbf{k} \chi_k(\eta) \tag{34}$$

leads to the operators of creation and annihilation with the commutation rules $[\hat{a}_\mathbf{k}, \hat{a}^+_\mathbf{k}] = 1$. Averaging over a vacuum state $\hat{a}_\mathbf{k}|0>=0$ gives:

$$<0|H|0> = -\frac{1}{2}M_p^2 a'^2 + \frac{1}{6}M_p^2 \Lambda a^4 + \frac{1}{2}\sum_{\mathbf{k}} a^2 \chi'^*_k \chi'_k + a^2 k^2 \chi^*_k \chi_k + a^4 m^2 \chi_k \chi_k. \tag{35}$$

The complex functions $\chi_k(\eta)$ satisfy the relations [35]:

$$\chi''_k + (k^2 + a^2 m^2)\chi_k + 2\frac{a'}{a}\chi'_k = 0,$$
$$a^2(\eta)(\chi_k \chi'^*_k - \chi^*_k \chi'_k) = i. \tag{36}$$

Eqs. (36) admit a formal WKB solution [35]:

$$\chi_k = \frac{\exp\left(-i\int_0^\eta W_k(\tau)d\tau\right)}{\sqrt{2}a(\eta)\sqrt{W_k(\eta)}}, \tag{37}$$

where the function $W_k(\eta)$ satisfies the equation

$$W''_k - \frac{3W'^2_k}{2W_k} - 2\left(k^2 + m^2 a^2 - \frac{a''}{a}\right)W_k + 2W_k^3 = 0. \tag{38}$$

Adiabatic approximation consists in setting

$$W_k(\eta) \approx \sqrt{k^2 + m^2 a(\eta)^2 - a''(\eta)/a(\eta)}.$$

Changing summation over $\mathbf{k}$ by integration

$$\sum_\mathbf{k} A_k \to \frac{4\pi}{(2\pi)^3}\int_0^{k_{max}} A_k k^2 dk, \tag{39}$$

and keeping the main terms at $k_{max}$, we come to

$$<0|H|0> = -\frac{1}{2}M_p^2 a'^2 + \frac{1}{6}M_p^2 \Lambda a^4 + \rho a^4 +$$
$$\frac{1}{2}\frac{4\pi}{(2\pi)^3}\left(\frac{k^4_{max}}{4} + \frac{k^2_{max}(m^2 a^4 + a'^2)}{4a^2}\right) = const. \tag{40}$$

Here, we have added `by hands" the term $\rho a^4$ corresponding to the dust matter satisfying $\rho a^3 = \rho_0 a_0^3$, where $a_0 = 1$ is the present-day value of the scale factor, $\rho_0$ is the present-day dust matter density. One may see, that the constant in FVT absorbs the leading part of the vacuum energy $\sim k_{max}^4$ during the whole evolution of universe. On the other hand, its $a$-dependence is similar to that of radiation density and does not relate to the contribution of $\Lambda$ having different $a$-dependence. In contrast, the unimodular gravity allows arbitrary cosmological constant [15], but that does not solve the vacuum energy problem at a fixed UV cut off of the comoving momentums $k_{max}$. For instance, if we introduce the cosmological constant compensating the vacuum energy at present, then the vacuum energy becomes large again at the time of the last scattering surface, i.e., at redshift $z = 1100$. That results from time-dependence of vacuum energy on the scale factor, whereas $\Lambda$ has not such a dependence. Actually, from (40) it follows $\rho_{vac} a^4 \sim k_{max}^4$, i.e. $\rho_{vac} \sim k_{max}^4/a^4$. Below, $\Lambda$ will equal zero, since, we will consider an alternative to it.

Let us introduce the parameter $S_0 = \dfrac{\kappa_{max}^2}{8\pi^2 M_p^2}$ [33,34], where $\kappa_{max}$ is a UV cut-off of the present-day physical momentums $\kappa_{max} = k_{max}/a_0$. Defining the constant $\Omega_m$ connected with the matter density $\dfrac{1}{2a_0}(a'^2)_0 \Omega_m M_p^2 = \rho a^3 = \rho_0 a_0^3$ and using Eq. (40) lead to

$$a'^2 = \frac{(S_0 - 1 + \Omega_m(1 - a/a_0))(a'^2)_0 + S_0^2 a_0^2 m^2(a_0^2 - a^2)}{S_0 a_0^2 a^{-2} - 1}. \tag{41}$$

First of all, one has to note that we handle a theory with the Big Rip occurring at $a = a_0\sqrt{S_0}$ due to the denominator of (41). That is a higher value of the momentums cut-off results in a longer life of the universe. There is no a theoretical upper bound on the UV cut-off, but the lower one corresponds to $S_0 = 1$ is $\kappa_{max} = 2\pi\sqrt{\dfrac{2}{N_{sc} + 2}} M_p$, where $N_{sc}$ is the number of scalar fields in theory [33], and the number two corresponds to a number of degrees of freedom for gravitational waves. At a bottom cut-off bound $S_0 < 1$, we would already be witnesses of the Big Rip.

It should be noted that the mass term is in fact $m^2 = \sum m_{bosons}^2 - m_{fermions}^2$ [33], i.e., the fermions contribute with opposite sign regarding the bosons one. The authors of [30] proved the theorem stating that adding any new bosons does not compensate all mass terms, and urged searching some new fermion families for this aim. In the light of the present work, the initial pre-condition of the theorem mentioned above [30] becomes more gentle, because there is no need in compensation of the main part associated with $k_{max}^4$ in the vacuum energy density. Here we assume for simplicity that all massive terms compensate each other in some wonderful way, and apply $m^2 = 0$ in the formula (41). Asymptotic of the solution of Eq. (41) in the vicinity of small-scale factors at $\eta \to -\infty$ is

$$a(\eta) \approx B\exp\left(\left(\frac{a'}{a}\right)_0 \sqrt{\frac{S_0+\Omega_m-1}{S_0}}\eta\right), \qquad (42)$$

where $B$ is some constant. We have in the cosmic time $dt = a(\eta)d\eta$ in the vicinity of $t=0$

$$a(t) \approx (\dot{a})_0 \sqrt{\frac{S_0+\Omega_m-1}{S_0}}\, t, \qquad (43)$$

where $\dot{a} = \dfrac{da}{dt}$. The deceleration parameter in terms of redshifts $z = a_0/a - 1$ is [34]

$$\begin{aligned} q(z) &= -\frac{\ddot{a}a}{\dot{a}^2} = -\left(\frac{a''a}{a'^2}-1\right) = \\ &\frac{q_0\left(\Omega_m(2q_0+\Omega_m-2)z^2 + 2\left(\Omega_m^2-3\Omega_m+2\right)z + (\Omega_m-2)^2\right)}{(\Omega_m+z(z+2)(2q_0+\Omega_m-2)-2)(\Omega_m+z(2q_0\Omega_m+\Omega_m-2)-2)}, \end{aligned} \qquad (44)$$

where $q_0 = \dfrac{-2+\Omega_m}{2(S_0-1)}$ is the present-day $z=0$ value of the deceleration parameter. It changes from the present-day negative value at small redshifts to zero at large $z$. If the dust matter content is considerable, $q$ can become positive in some interval. It is interesting that $q(z)$ in the VFD model is weakly sensitive to a dust matter content as we see from Fig. 4.

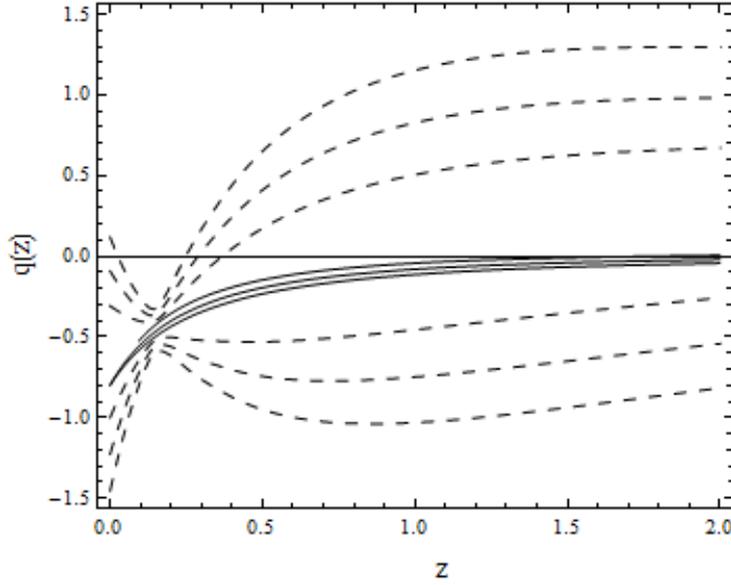

Fig. 4. Dependence of the deceleration parameter on redshift (bold) at $\Omega_m = 0.7, 0.27, 0.05$ put on the $1\sigma$, $2\sigma$, $3\sigma$ error channels (dashed) of the reconstruction of the deceleration parameter [36] from the 115 SN Ia data.

One may assume that the late time universe acceleration results from the residual vacuum fluctuations of a scalar field. At least one scalar field is already discovered, that is the Higgs boson. Besides, it was shown [33] that gravitational waves should also produce the analogous effect. A linear Milne-like expansion precedes this accelerated stage of universe evolution. It is interesting that the Milne-like universes again retain the great attention. It was shown that the primordial nucleosynthesis is concordant with the observational data within the framework of such models [37]. The other cosmological tests are also under discussion [38-44].

**5. Conclusion.** On the one hand, the results of the paper could be considered from the abstract point of view as the possibility of a gravity theory admitting an arbitrary choice of the energy density reference level. We have introduced the surface of the constraints $\partial_i \mathcal{H} = 0$ and $\mathcal{P}_i = 0$ instead of the surface $\mathcal{H} = 0$ and $\mathcal{P}_i = 0$, and have found the Hamiltonian, which governs a system evolution along the former surface. The FVT is completely self-consistent in terms of the theory of constrained systems [19,20]. The price is that time and space are not considered as a single $R^4$ manifold.

On the other hand, the remarkable property of the theory is that the main part of vacuum energy does not influence the universe evolution. However, there remains an open question regarding the contribution of masses of particles into the vacuum energy. In any case, FVT is a strong argument for the VFD model [34], which predicts accelerated universe expansion at $z \leq 1$, and the Milne-like universe at $z \geq 1$. From the general point of view, the possibility to chose an arbitrary energy level in FVT seems analogous to that in nonrelativistic physics. After the compensation of the main part of vacuum energy, the theory becomes looking as GR in the conformal time gauge except for residual vacuum energy influencing the cosmological evolution.

**Acknowledgments.** This work is performed in a frame of the program of the scientific research of the Ministry of education of the Republic of Belarus "Convergence-2020", subprogram 2.1.07 "Microworld, plasma and universe."

**Information about the authors**

**Cherkas Sergey Leonidovich** – Ph. D. (Physics and Mathematics), Senior scientist, Institute for Nuclear Problems, Belarus State University (11, Bobruiskays Str., 220050, Minsk, Republic of Belarus). E-mail: cherkas@inp.bsu.by

**Kalashnikov Vladimir Leonidovich** – Ph. D. (Physics and Mathematics), Senior research fellow, Institute of Photonics, Vienna University of Technology, (27/387, Gusshausstrasse, A-1040, Vienna, Austria). E-mail: vladimir.kalashnikov@tuwien.ac.at

**С. Л. Черкас [1], В. Л. Калашников[2]**

[1]*НИИ Ядерных проблем при Белорусском государственном университете, Минск, Беларусь*
[2]*Венский технический университет, Вена, Австрия*

# ТЕОРИЯ ГРАВИТАЦИИ С ПРОИЗВОЛЬНЫМ УРОВНЕМ ОТСЧЕТА ПЛОТНОСТИ ЭНЕРГИИ

Предложена пятивекторная теория гравитации, в которой уровень отсчета плотности энергии может быть выбран произвольно. Теория сформулирована, как система со связями, в которой множители Лагранжа принадлежат некоторому ограниченному классу векторных полей, в отличие от Общей Теории относительности, где множители Лагранжа могут быть заданы произвольно. Следствием теории является утверждение, что основная часть вакуумной плотности энергии не влияет на расширение вселенной, в то время как оставшаяся часть приводит к закону расширения, близкому к линейному, как у вселенной Миллна.

*Ключевые слова:* пятивекторная теория гравитации, Гамильтоновы системы со связями, вакуумная энергия, квантовая космология, космология Милна